\documentclass[aps,nofootinbib,floatfix,showpacs,preprintnumbers,twocolumn,superscriptaddress]{revtex4}
\usepackage{graphicx}
\usepackage{bm}
\usepackage{mathrsfs}
\usepackage{float}
\usepackage{tabularx} 
\usepackage{amsmath}
\usepackage{epstopdf}
\usepackage{xcolor}
\usepackage{hyperref}
\usepackage{ulem}
\usepackage{natbib}
\input epsf

\begin{document}

\title{First germanium-based constraints on sub-MeV Dark Matter \\ with the EDELWEISS experiment}
\author{Q. Arnaud}\email{q.arnaud@ipnl.in2p3.fr}
\affiliation{Univ Lyon, Universit\'e Lyon 1, CNRS/IN2P3, IP2I-Lyon, F-69622, Villeurbanne, France}
\author{E. Armengaud}
\affiliation{IRFU, CEA, Universit\'{e} Paris-Saclay, F-91191 Gif-sur-Yvette, France}
\author{C. Augier}
\affiliation{Univ Lyon, Universit\'e Lyon 1, CNRS/IN2P3, IP2I-Lyon, F-69622, Villeurbanne, France}
\author{A.~Beno\^{i}t}
\affiliation{Institut N\'{e}el, CNRS/UJF, 25 rue des Martyrs, BP 166, 38042 Grenoble, France}
\author{L.~Berg\'{e}}
\affiliation{Universit\'{e} Paris-Saclay, CNRS/IN2P3, IJCLab, 91405 Orsay, France}
\author{J.~Billard}
\affiliation{Univ Lyon, Universit\'e Lyon 1, CNRS/IN2P3, IP2I-Lyon, F-69622, Villeurbanne, France}
\author{A.~Broniatowski}
\affiliation{Universit\'{e} Paris-Saclay, CNRS/IN2P3, IJCLab, 91405 Orsay, France}
\author{P.~Camus}
\affiliation{Institut N\'{e}el, CNRS/UJF, 25 rue des Martyrs, BP 166, 38042 Grenoble, France}
\author{A.~Cazes}
\affiliation{Univ Lyon, Universit\'e Lyon 1, CNRS/IN2P3, IP2I-Lyon, F-69622, Villeurbanne, France}
\author{M.~Chapellier}
\affiliation{Universit\'{e} Paris-Saclay, CNRS/IN2P3, IJCLab, 91405 Orsay, France}
\author{F.~Charlieux}
\affiliation{Univ Lyon, Universit\'e Lyon 1, CNRS/IN2P3, IP2I-Lyon, F-69622, Villeurbanne, France}
\author{M. De~J\'{e}sus}
\affiliation{Univ Lyon, Universit\'e Lyon 1, CNRS/IN2P3, IP2I-Lyon, F-69622, Villeurbanne, France}
\author{L.~Dumoulin}
\affiliation{Universit\'{e} Paris-Saclay, CNRS/IN2P3, IJCLab, 91405 Orsay, France}
\author{K.~Eitel}
\affiliation{Karlsruher Institut f\"{u}r Technologie, Institut f\"{u}r Kernphysik, Postfach 3640, 76021 Karlsruhe, Germany}
\author{E.~Elkhoury}
\affiliation{Univ Lyon, Universit\'e Lyon 1, CNRS/IN2P3, IP2I-Lyon, F-69622, Villeurbanne, France}
\author{J.-B.~Fillipini}
\affiliation{Univ Lyon, Universit\'e Lyon 1, CNRS/IN2P3, IP2I-Lyon, F-69622, Villeurbanne, France}
\author{D.~Filosofov}
\affiliation{JINR, Laboratory of Nuclear Problems, Joliot-Curie 6, 141980 Dubna, Moscow Region, Russian Federation}
\author{J.~Gascon}
\affiliation{Univ Lyon, Universit\'e Lyon 1, CNRS/IN2P3, IP2I-Lyon, F-69622, Villeurbanne, France}
\author{A.~Giuliani}
\affiliation{Universit\'{e} Paris-Saclay, CNRS/IN2P3, IJCLab, 91405 Orsay, France}
\author{M.~Gros}
\affiliation{IRFU, CEA, Universit\'{e} Paris-Saclay, F-91191 Gif-sur-Yvette, France}
\author{Y.~Jin}
\affiliation{C2N, CNRS, Univ.  Paris-Sud, Univ.  Paris-Saclay, 91120 Palaiseau, France}
\author{A.~Juillard}
\affiliation{Univ Lyon, Universit\'e Lyon 1, CNRS/IN2P3, IP2I-Lyon, F-69622, Villeurbanne, France}
\author{M.~Kleifges}
\affiliation{Karlsruher Institut f\"{u}r Technologie, Institut f\"{u}r Prozessdatenverarbeitung und Elektronik, Postfach 3640, 76021 Karlsruhe, Germany}
\author{H.~Lattaud}
\affiliation{Univ Lyon, Universit\'e Lyon 1, CNRS/IN2P3, IP2I-Lyon, F-69622, Villeurbanne, France}
\author{S.~Marnieros}
\affiliation{Universit\'{e} Paris-Saclay, CNRS/IN2P3, IJCLab, 91405 Orsay, France}
\author{D.~Misiak}
\affiliation{Univ Lyon, Universit\'e Lyon 1, CNRS/IN2P3, IP2I-Lyon, F-69622, Villeurbanne, France}
\author{X.-F.~Navick}
\affiliation{IRFU, CEA, Universit\'{e} Paris-Saclay, F-91191 Gif-sur-Yvette, France}
\author{C.~Nones}
\affiliation{IRFU, CEA, Universit\'{e} Paris-Saclay, F-91191 Gif-sur-Yvette, France}
\author{E.~Olivieri}
\affiliation{Universit\'{e} Paris-Saclay, CNRS/IN2P3, IJCLab, 91405 Orsay, France}
\author{C.~Oriol}
\affiliation{Universit\'{e} Paris-Saclay, CNRS/IN2P3, IJCLab, 91405 Orsay, France}
\author{P.~Pari}
\affiliation{IRAMIS, CEA, Universit\'{e} Paris-Saclay, F-91191 Gif-sur-Yvette, France}
\author{B.~Paul}
\affiliation{IRFU, CEA, Universit\'{e} Paris-Saclay, F-91191 Gif-sur-Yvette, France}
\author{D.~Poda}
\affiliation{Universit\'{e} Paris-Saclay, CNRS/IN2P3, IJCLab, 91405 Orsay, France}
\author{S.~Rozov}
\affiliation{JINR, Laboratory of Nuclear Problems, Joliot-Curie 6, 141980 Dubna, Moscow Region, Russian Federation}
\author{T.~Salagnac}
\affiliation{Univ Lyon, Universit\'e Lyon 1, CNRS/IN2P3, IP2I-Lyon, F-69622, Villeurbanne, France}
\author{V.~Sanglard}
\affiliation{Univ Lyon, Universit\'e Lyon 1, CNRS/IN2P3, IP2I-Lyon, F-69622, Villeurbanne, France}
\author{B.~Siebenborn}
\affiliation{Karlsruher Institut f\"{u}r Technologie, Institut f\"{u}r Kernphysik, Postfach 3640, 76021 Karlsruhe, Germany}
\author{L.~Vagneron}
\affiliation{Univ Lyon, Universit\'e Lyon 1, CNRS/IN2P3, IP2I-Lyon, F-69622, Villeurbanne, France}
\author{M.~Weber}
\affiliation{Karlsruher Institut f\"{u}r Technologie, Institut f\"{u}r Prozessdatenverarbeitung und Elektronik, Postfach 3640, 76021 Karlsruhe, Germany}
\author{E.~Yakushev}
\affiliation{JINR, Laboratory of Nuclear Problems, Joliot-Curie 6, 141980 Dubna, Moscow Region, Russian Federation}
\author{A.~Zolotarova}
\affiliation{Universit\'{e} Paris-Saclay, CNRS/IN2P3, IJCLab, 91405 Orsay, France}

\collaboration{EDELWEISS Collaboration} 
\noaffiliation

%\vspace{.2in}

\date{\today}

\smallskip
\begin{abstract}The EDELWEISS collaboration has performed a search for Dark Matter (DM) particles interacting with electrons using a 33.4~g Ge cryogenic detector operated underground at the LSM.  A charge resolution of 0.53~electron-hole pairs (RMS) has been achieved using the Neganov-Trofimov-Luke amplification with a bias of 78~V.
We set the first Ge-based constraints on sub-MeV/c$^{2}$ DM particles interacting with electrons, as well as on dark photons down to 1 eV/c$^2$. These are competitive with other searches. In particular, new limits are set on the kinetic mixing of dark photon DM in a so far unconstrained parameter space region in the 6 to 9 eV/c$^2$ mass range. These results demonstrate the high relevance of cryogenic Ge detectors for the search of DM interactions producing eV-scale electron signals. %By using the most massive semiconductor crystal used for these searches, this work paves the way to the development of kg-size detector arrays.
\end{abstract}

\maketitle

%\section{Introduction}

Direct-detection experiments are progressing rapidly in the search of nuclear scattering events due to Weakly Interacting Massive Particles on the GeV/c$^2$ to TeV/c$^2$ mass scale~\cite{xenon1t,lux,pandax,DarksideWIMPs}.
However, there are compelling models that motivate to extend direct searches to Dark Matter (DM) particles in the eV/c$^2$ to MeV/c$^2$ range, where the signal would be an  electron recoil arising either from the absorption of a dark photon (bosonic DM)~\cite{DPreference,DarkPhotonAbsorption}, or the elastic scattering of a dark fermion~\cite{Essig2016}. For these searches -- requiring kg-scale detectors with $\sim1$~eV detection thresholds to fully cover benchmark models~\cite{com} -- semiconductor detectors are uniquely positioned due to their band-gap energies an order of magnitude lower than the ionization potential of xenon-based detectors~\cite{Xenon10-DP}. 

Recent progress has been made with silicon-based gram-scale devices, using CCDs~\cite{SENSEI@MINOS,DAMIC-2019} and cryogenic detectors~\cite{CDMS-2019-erratum} now sensitive to single electron-hole pairs. Efforts are ongoing to reduce dark currents and radioactive background to the levels required for scaling up to more massive arrays. In this context, phonon-mediated germanium detectors offer an attractive alternative. The smaller band-gap energy of Ge relative to Si ($E_{g}=0.67~\mathrm{eV}$ vs. $1.11~\mathrm{eV}$~\cite{refgap1,refgap2}) naturally yields an increased sensitivity to lighter DM particles. In addition, the difference in composition paves the way to a better understanding of the origin of the background observed in semiconductor detectors at this new eV-scale frontier. 

In phonon-mediated cryogenic detectors, the drift of $N$ electron-hole pairs across a voltage difference $\Delta V$ produces additional phonons whose energy $E_{\mathrm{NTL}}=N\Delta V$ (in eV) adds up to the initial recoil energy. This effect called Neganov-Trofimov-Luke (NTL)~\cite{Neganov,Luke} essentially turns a cryogenic calorimeter (operated at $\Delta V$=0~V) into a charge amplifier of mean gain $\left<g\right>=(1+\Delta V/\epsilon)$,  where $\epsilon=3.0$~eV ($3.8$~eV) is the mean ionization energy in Ge~(Si)~\cite{Ge3eV} for electron recoils. %For a given readout noise, the resolution improves as $1/\left<g\right>$.

Recently, the EDELWEISS collaboration achieved a 17.7~eV phonon baseline resolution (RMS) with a 33.4~g Ge bolometer operated above-ground~\cite{RED20}. To reach sub-electron-hole pair resolution, a similar detector was equipped with electrodes to take advantage of the expected $1/\left<g\right>$ improvement of the charge resolution with applied voltage. 
In this letter, we exploit the resulting sensitivity to energy deposits as low as the band-gap energy to set competitive constraints on sub-MeV/c$^2$ DM particles interacting with electrons, as well as on dark photons down to 1 eV/c$^2$. %New limits are set on hidden-photon DM in the range 1.2--30~eV/c$^2$ mass range, improving upon existing solar and direct detection constraints~\cite{SENSEI@MINOS,CDMS-2019-erratum,DAMIC-2019,Xenon10-DP,Darkside,Solar2,Solar3}.

The DM search was performed at the Laboratoire Souterrain de Modane (France) with a detector consisting of a 33.4~g cylindrical high-purity Ge crystal ($\oslash$20$\times$20~mm). Aluminum electrodes are lithographed on both planar surfaces in a grid scheme, except for the outer edges where they are shaped as concentric guard rings. 
A $2\times2$~$\mathrm{mm^2}$ area was left empty at the center of one face to allow for the direct gluing of a Ge neutron-transmutation-doped (NTD)~\cite{NTD} thermal sensor on the crystal. The center and guard electrodes on the same side are biased to the same voltage, but the associated ionization channels are read out separately. The data acquisition system and readout electronics are the same as in~\cite{edwtech}.
%, modified to withstand an electrode bias range of $\pm70$~V~\cite{BBHV}. 
The data from the phonon and ionization channels were digitized at a frequency of 100~kHz, filtered, averaged and continuously stored on disk with a digitization rate of 500~Hz. For the DM search, only the phonon channel is used.

\begin{figure}[t]
\includegraphics[width=0.5\textwidth,angle=0]{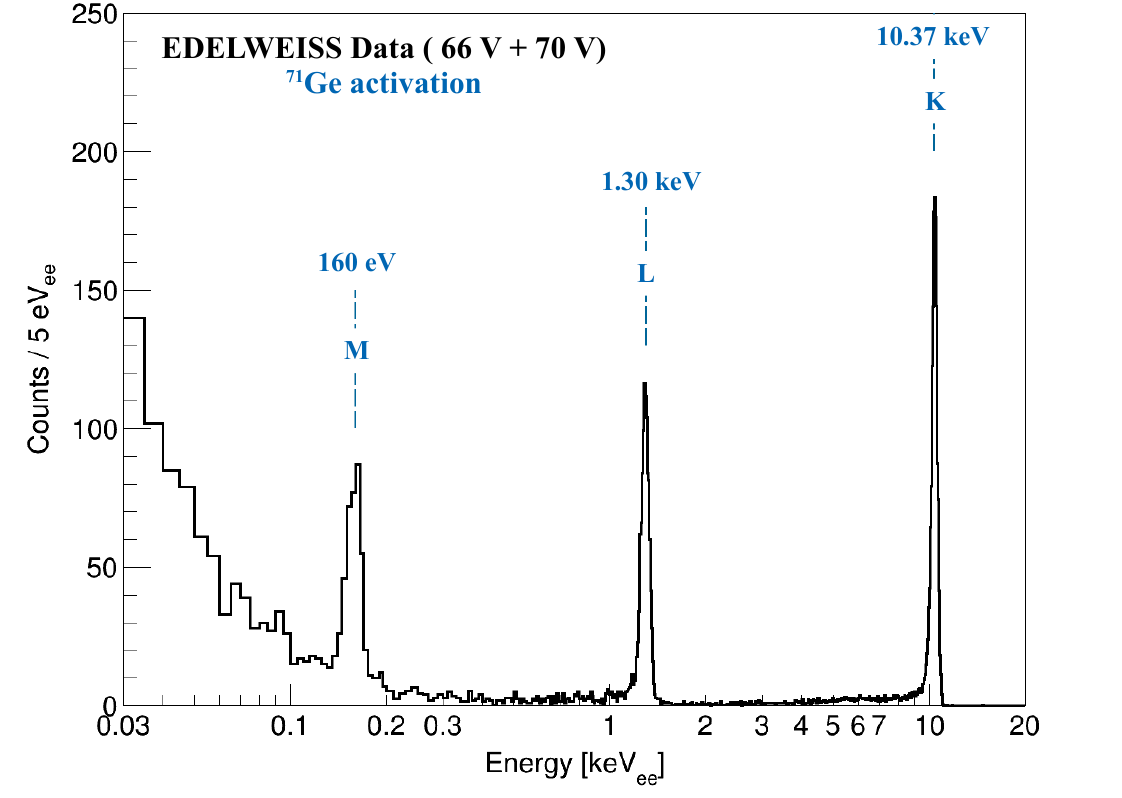}
\caption{Energy spectrum recorded with a bias of 66~V and 70~V following the ${}^{71}\mathrm{Ge}$ activation of the detector. 
} 
\label{fig:spectrum}
\end{figure}

%\textcolor{red}{using epoxy (EPO-TEK 301-2). The crystal is held with 3 PTFE clamps on the detector side hosting the NTD and with 3 sapphire balls on the other side. The NTD sensor was current-biased at $\pm$1.5~nA to a resistance value of 1.5 M$\Omega$.}
%%% Do we have published proof for the following???
%A square meshing with a 500~$\mathrm{\mu m}$ pitch was adopted for the grid electrode scheme to minimize Al coverage ($\sim$4\%) and hence phonon trapping, in an attempt to maximize the sensitivity of the NTD to athermal phonons. 
%%% We don't have published proof that the next item is true too!
%In this spirit and to further reduce risk of leakage with the operation of the detector at a high voltage, the lateral surfaces were left unequipped with electrodes. 
%The crystal is held with 3 PTFE clamps on detector side hosting the NTD and with 3 sapphire balls on the other side.
 %%% ...the rest is not necessarily true!
%%% to maximize the uniformity electric field in the crystal. 
The detector was maintained at a regulated temperature of either 20.0 or 20.7 mK between January and October 2019. 
Most of that period was devoted to detector studies and calibrations. Prior to its installation in the cryostat, the detector was uniformly activated using a neutron AmBe source. The produced short-lived isotope $^{71}$Ge decays by electron capture in the K, L and M shells, emitting characteristic lines at 10.37, 1.30 and 0.16 keV, respectively. The activation lines are locally absorbed, thus providing very good probes of the detector response to a DM signal uniformly distributed inside the detector volume. These are clearly visible in Fig.~\ref{fig:spectrum}, which shows the energy spectrum -- in units of eV-electron-equivalent (eV$_{\mathrm{ee}}$) -- from calibration data recorded in January at biases of 66~V and 70~V. The measured L/K and M/L yields of $0.110\pm0.008$ and $0.158\pm0.020$ are compatible with existing measurements~\cite{KLMratios1,KLMratios2}.
The resolution on the 160~eV peak of $\sigma=8~\mathrm{eV_{ee}}$ is consistent with the Fano factor $\mathrm{F}=0.15$ expected for Ge at low energy~\cite{FanoGe}. The precision on the K-line position is better than 0.1\%. By varying the bias from 0~V to 81~V, the non-linearity of the heat sensor signal was measured to be less than 5\% over three orders of magnitude, with a 2\% uncertainty extrapolation down to zero energy.

The K and L peaks are accompanied by a low-energy tail of events. On the basis of the corresponding signals observed on the center and guard electrodes, these tails are ascribed to incomplete charge collection for events near the cylindrical surfaces. To prevent charge build-up that would otherwise worsen the collection performance, the detector was regularly grounded for periods of 2--10~h while being exposed to a strong $^{60}$Co source. This regeneration procedure allows us to neutralize residual fields induced by the accumulation of trapped charges~\cite{edwtech}. The tail represents 19\% of the K line events above $1.5~\mathrm{ keV_{ee}}$ in Fig.~\ref{fig:spectrum}. No significant increase of that tail is observed in the days following a regeneration, with an upper limit of +1\% per day.  
%This tail increases by less than 1\%/day

%This temporary noise increase appears to depend on charge build-up in the crystal, as the magnitude of the effect varies with the exposure, the event rate, and the applied voltage.

%\section{Data processing}

%As in Ref.~\cite{RED20}, the amplitude of the pulses are evaluated by minimizing a $\chi^2_k$ function defined in the frequency domain.

%\section{Calibration and efficiency determination}
%\label{sec:cal}

%\section{Search Data}
A bias of up to 81~V could be applied  without heating up the detector.
Ramping up the bias produces an additional noise on the phonon channel.
Most of it ebbs away after a period of 12--72~h, after which the baseline resolution at 78~V is typically 10\% above its value at $\sim$0~V, once the NTL amplification is taken into account. 
Attempts to use the ``pre-biasing" method~\cite{CDMS-2019-erratum} did not significantly reduce this period.

The study of the detector performance and stability led to the choice of a bias of 78~V for a DM search involving electron recoils. 
A continuous sequence of runs at 78~V from April 1$^{st}$ to 7$^{th}$ were set aside for this search. The baseline resolution derived from random trigger samples was studied hour by hour. The first three days were discarded as the baseline resolution reached its plateau only at the end of that period.
The remaining 89~h of data were separated into a blind sample of 58~h sandwiched between a non-blind sample of 21~h plus 9~h of data. The stability of the energy scale was monitored day by day using the K-line peak. The average baseline energy resolution in the non-blind sample is 1.63~eV$_{\mathrm{ee}}$ (0.54 electron-hole pairs), corresponding to a phonon resolution of 44~eV, once the NTL gain of 27 is considered. The average baseline resolution in the blind sample is 3\% better (1.58 eV$_{\mathrm{ee}}$, or 0.53 pairs).
Three days after the data taking described above, the detector was exposed again to a strong AmBe source for 15~h, in order to reactivate it and confirm the stability of the detector response with high statistics.

The data processing -- based on an optimal matching filter approach -- is essentially the same as in~\cite{RED20} and uses the numerical procedure described in~\cite{trigger}.
An iterative search for pulses in the filtered data stream is performed using a decreasing energy ordering rule. After the pulse with the largest amplitude is found, a time trace of $\Delta t=2.048~\mathrm{s}$ is allocated. This period is excluded from the search in the next iteration, proceeding downward in amplitude. The procedure stops when there is no time interval greater than $\Delta t$ left in the stream. Thus, there is no trigger threshold set in energy and the trigger rate is driven by the choice of $\Delta t$, not by the physical event rate. The energy dependence of the dead time induced by this procedure is fully taken into account in the evaluation of the trigger efficiency using a pulse simulation described below.

The pulse amplitudes are evaluated by minimizing a $\chi^2_k$ function in the frequency domain.%~\cite{RED20}.
%\begin{equation}
%    \chi^2_k(a,t_0)=\sum_i \frac{|{\overline{v}(f_i)-a\tilde{s_k}(f_i)e^{-j2\pi t_0 f_i}}|^2}{J(f_i)}
%\end{equation}
%where $J(f_i)$ is the power spectral density (PSD), ${\overline{v}(f_i)}$ is the Fourier transform of the trace, $t_0$ is the starting time of the pulse and $a$ is the amplitude of the signal template $s_k$ normalized to unity. 
%Similarly than in Ref.~\cite{RED20},
The subscript $k$ designates the so-called ``normal'' and ``fast'' categories of events, each corresponding to a different pulse template. Normal events refer to particle interactions occurring in the Ge target crystal. For this category, we use a template based on 10.37 keV event pulses which are characterized by a rise time of $\sim7~\mathrm{ms}$. Fast events stand out with a considerably shorter rise time ($<1$~ms), compatible with interactions occurring directly in the NTD. The data selection is based on the values of $\chi^2_{normal}$ and on the difference $\Delta \chi^2=\chi^2_{normal}-\chi^2_{fast}$ whereas pulse amplitude estimation is based on the normal template only.

\begin{figure}[t]
\includegraphics[width=0.5\textwidth,angle=0]{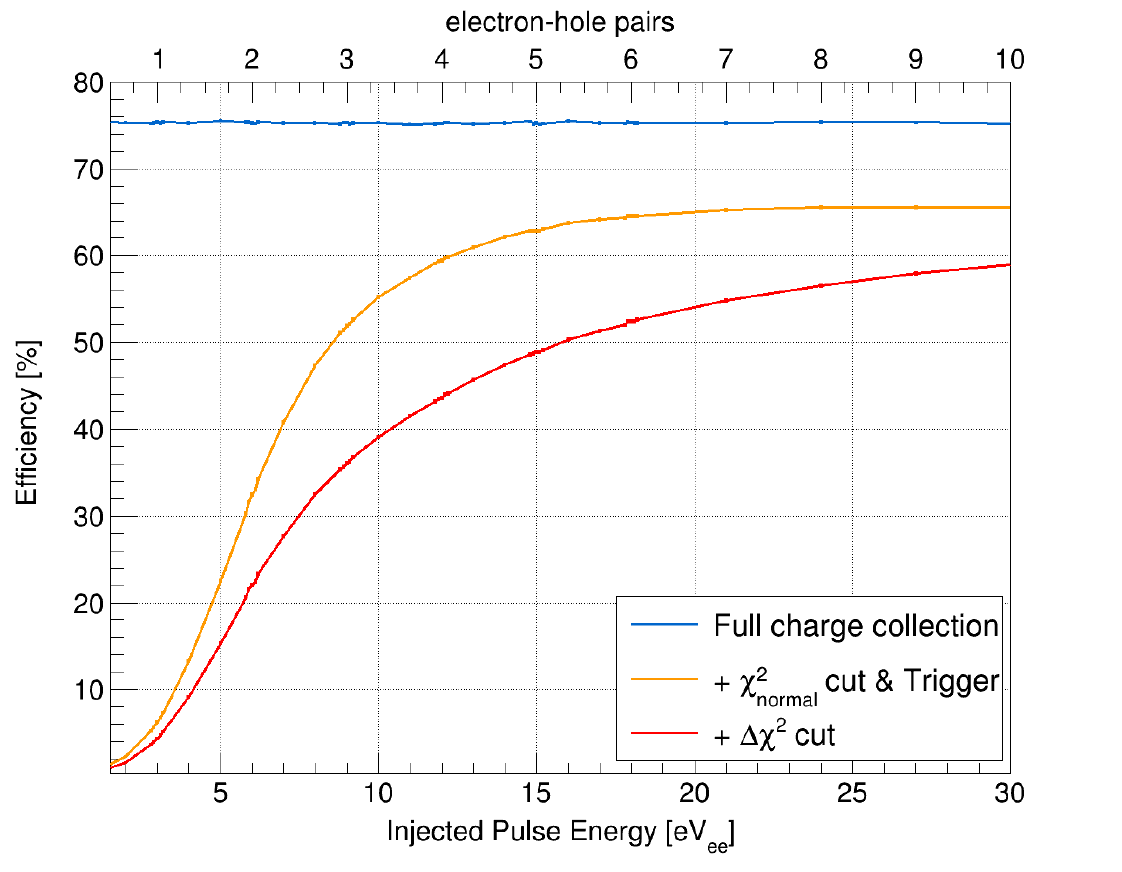}
\caption{Signal efficiency measured as the fraction of pulses injected in the data stream that pass successive selection criteria, as a function of the input phonon energy.} 
\label{fig:efficiency}
\end{figure}

\begin{figure}[t]
\includegraphics[width=0.5\textwidth,angle=0]{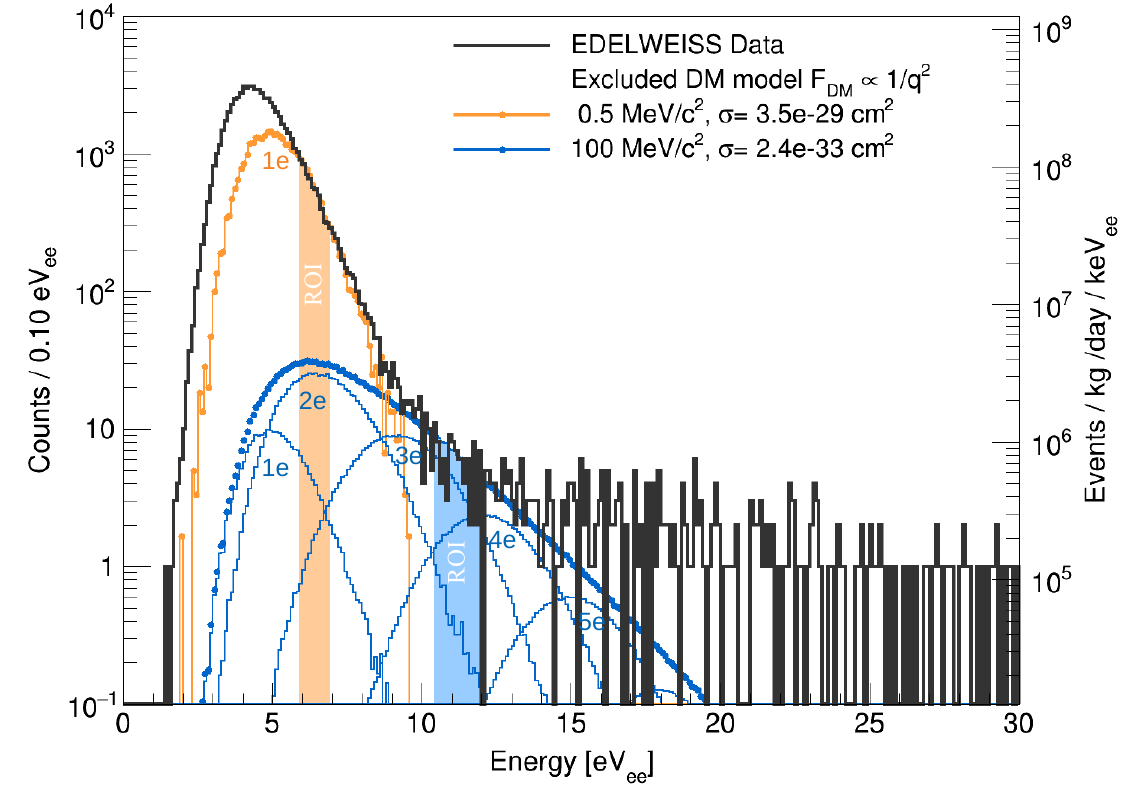}
\caption{
Energy spectrum of the events selected for the DM search (black). The thick blue (orange) histogram is the simulation of the signal excluded at 90\%~C.L. for a DM particle with a mass of 10 (0.5) MeV/c$^2$, and $F_{DM}=1/q^2$. 
The thin-line histograms of the same color represent the individual contributions of 1 to 5 electron-hole pairs.
The corresponding ROIs used to set the upper limits are shown as shaded intervals using the same color code.
} 
\label{fig:data}
\end{figure}

The trigger and cut efficiencies were determined using a complete signal simulation procedure~\cite{RED20}.
%The data analysis uses a complete signal simulation to determine trigger and cut efficiencies~\cite{RED20}.
Pulses of known energy are injected at random times throughout the entire real data streams at a rate of $\sim 0.02~\mathrm{Hz}$ in order not to increase the deadtime by more than $\sim1\%$.
Each simulated pulse corresponds to a trace randomly chosen among a selection of K-line events, scaled to the desired fraction of 10.37 keV and added to the data stream. The set of preselected traces consists of 858 K-line events with energies between 1.5 and 11~$\mathrm{keV_{ee}}$, recorded at 78~V after post-search activation. The effect of incomplete charge collection on the signal detection efficiency is conservatively accounted for by ascribing a 0\% survival probability to simulated pulses from traces of K-line events with an energy $470~\mathrm{eV_{ee}}$ ($\sim3\,\sigma$) away from the 10.37~$\mathrm{keV_{ee}}$ peak. This results in a $25\%$ efficiency loss\footnote{This is larger than the 19\% tail fraction from Fig.~\ref{fig:spectrum} due to the reduced  K-line event rate (3~vs.~15~mHz) in the post-search activation sample compared to a constant Compton background.} visible in~Fig.~\ref{fig:efficiency} showing the trigger and analysis cut efficiencies as a function of the injected pulse energy. The fraction of simulated events surviving the reconstruction procedure and the selection cut on the value of $\chi^2_{normal}$ are shown as orange line. The plateau reduction from 75\% to 65\% is due in large part to losses due to reset periods required for the operation of the charge readout. 
The decrease of efficiency at low energy is due to the trigger algorithm bias toward high-energy events. The cut on $\Delta\chi^2$ further reduces the plateau to 59\% at 30~eV$_{\mathrm{ee}}$.
After all cuts the efficiency for simulated single (double) electron-hole pair events of 3 eV$_{\mathrm{ee}}$ (6 eV$_{\mathrm{ee}}$) is 4\% (22\%).
Fig.~\ref{fig:data} shows the energy spectrum of the selected events in the 58~hours of DM search. The efficiency-corrected rate at 25 eV$_{\mathrm{ee}}$ corresponds to $1.6\times10^{5}\,\mathrm{events/kg/day/keV_{ee}}$. Comparison of the spectra recorded at different biases suggests that most of the rate observed above 25 eV$_{\mathrm{ee}}$ at 78~V (0.68~keV total phonon energy) corresponds to events not affected by the NTL amplification. The origin of these so-called ``heat-only'' events is still under investigation\footnote{The rate observed at $25~\mathrm{eV}_{ee}$ would correspond to a background of $6\times10^{3}$ events/kg/day/keV at 0.68 keV if these are assumed to have no associated charge. This is a factor 3 below the background observed above ground at this energy in~\protect{\cite{RED20}}.}.
Although it limits the detection of electron recoils above $\sim$10~eV$_{\mathrm{ee}}$, the main limitation for signals associated to fully-collected 1 to 3 electron-hole pairs is the rapid rise of the spectrum at low energy.
In Fig.~\ref{fig:data}, we show the contributions of $N=[1,...,5]$ electron-hole pair events obtained from the pulse simulation for models described below. For $N>1$, the reconstructed energy spectra associated to $N$-pair events peak at $N\times\epsilon$~eV$_{\mathrm{ee}}$. However, the spectrum associated to single-pair events is biased towards higher energy as only those with reconstructed energies above $\sim3$~eV are selected by the trigger algorithm.
The detector resolution is not sufficient to unambiguously disentangle single-pair from noise-triggered events. 
%\textcolor{red}{fully-collected single-pair events from fluctuations of the noise on the phonon read-out channel or partially collected charges.} 
It is however able to provide an upper bound on single-pair (or $N$-pair events), and more generally to the DM signals discussed below.

%\section{Results}
\begin{figure}[htbp]
\includegraphics[width=0.5\textwidth,angle=0]{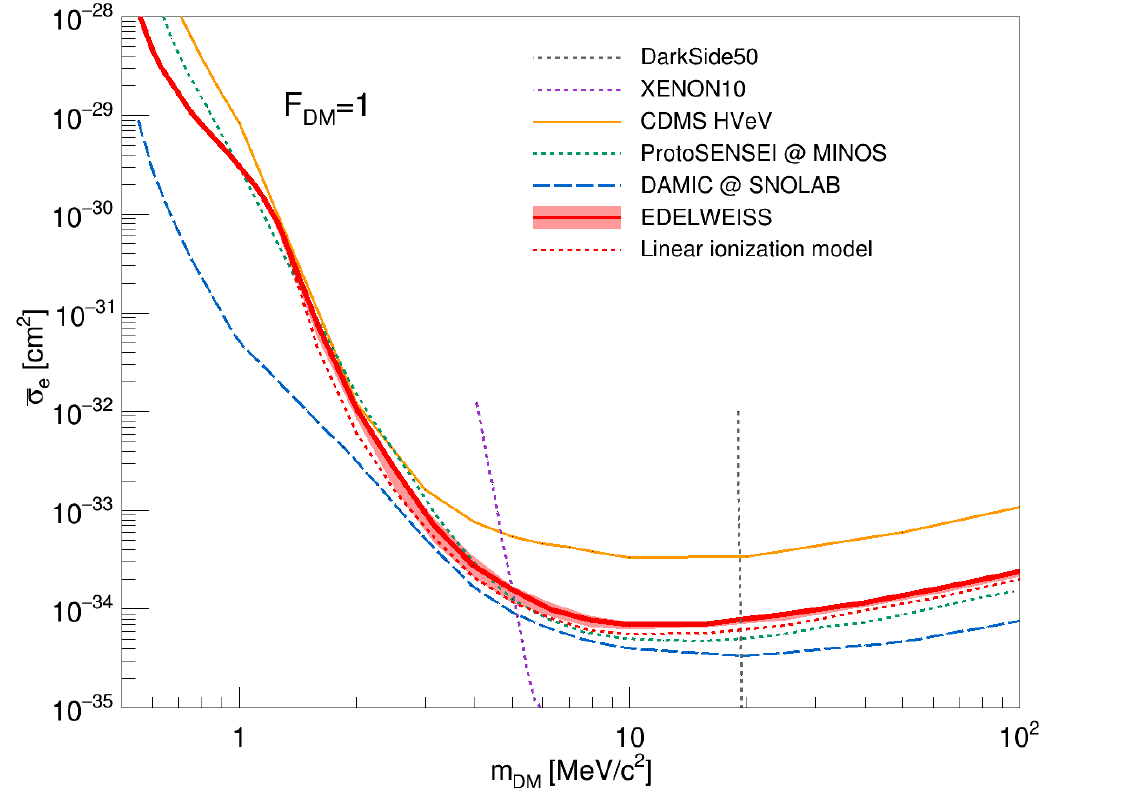}
\includegraphics[width=0.5\textwidth,angle=0]{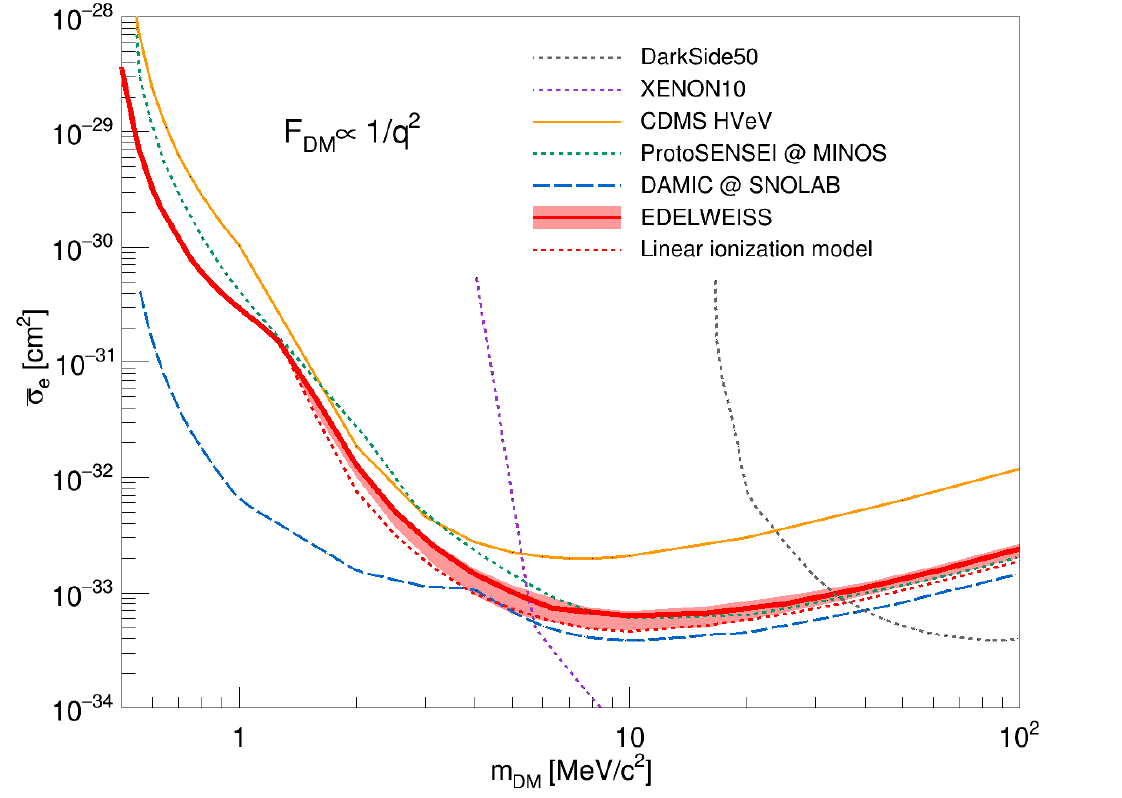}
\includegraphics[width=0.5\textwidth,angle=0]{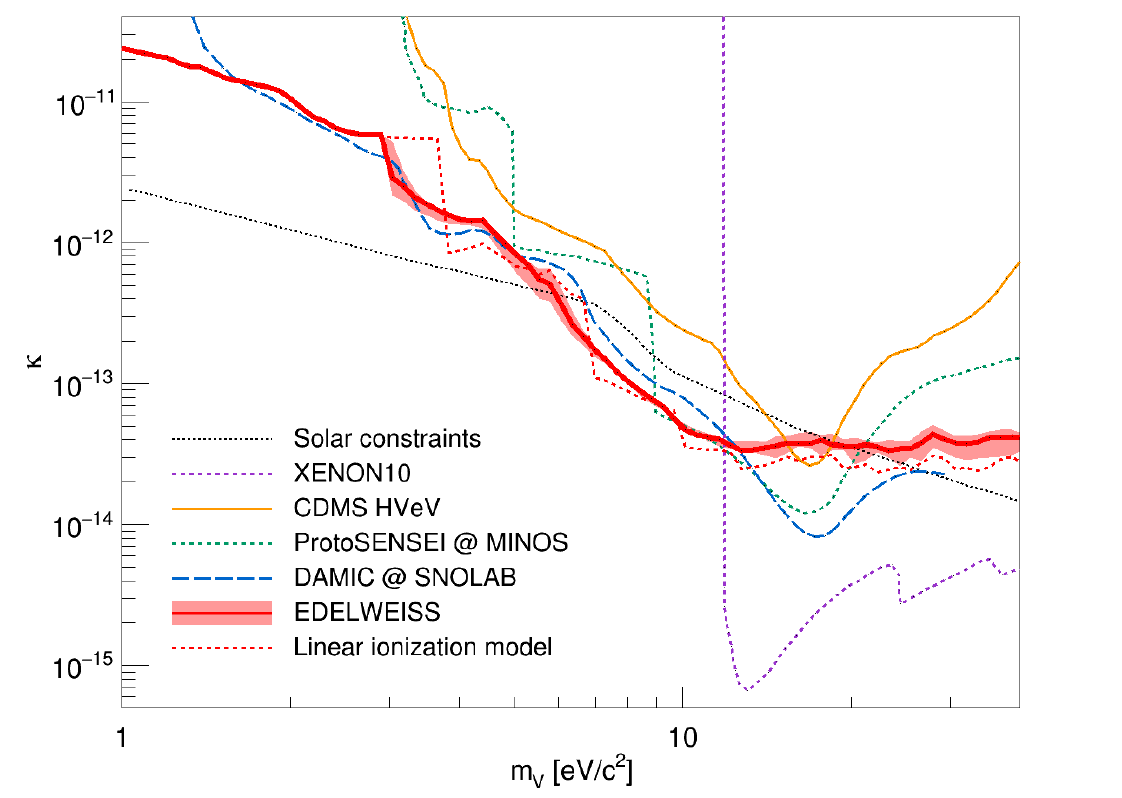}
\caption{
90\%~C.L. upper limit on the cross section for the scattering of DM particles on electrons, assuming a heavy (top panel) or light (middle panel) mediator. Bottom: 90\%~C.L. upper limit on the kinetic mixing $\kappa$ of a dark photon. The results from the present work are shown as the red line. The shaded red band and dotted red line represent alternative charge distribution models (see text). Also shown are constraints from other direct detection experiments~\protect{\cite{SENSEI@MINOS,CDMS-2019-erratum,DAMIC-2019,Xenon10-DP,Darkside,Essig2016}}, and solar constraints~\protect{\cite{Solar2,Solar3}}.
} 
\label{fig:results}
\end{figure}
%\textcolor{red}{The differential rate of electron recoils induced by DM scattering calculated according to~\cite{Essig2016} as follows:}
The DM-electron scattering rate as a function of the energy transfered to the electron $E_e$ is given by~\cite{Essig2016}:
\begin{equation}
    \frac{dN}{{dE}_{e}}\propto {\overline{\sigma}}_{e} \int \frac{d q}{q^2} \eta(E_{e},q,m_{\chi})|F_{DM}(q)|^2 |f_{c}(q,E_{e})|^2,
    \label{eq:ERDM}
\end{equation}
where ${\overline{\sigma}}_{e}$ is a reference cross section for free electron scattering. 
The term $\eta$ encapsulates DM halo physics and is calculated assuming a local DM density of $\rho=0.3$~$\mathrm{GeV}/\mathrm{c^2}/\mathrm{cm^3}$, a galactic escape velocity $\mathrm{v_{esc}}$= 544~km/s and an asymptotic circular velocity $\mathrm{v_{0}}$= 220~km/s~\cite{standardassumptions1,standardassumptions2}. 
The momentum-transfer $q$ dependence of the interaction is described by the form factor $F_{DM}$. 
The crystal form factor $f_{c}$ is related to the probability that a momentum transfer $q$ yields an electron transition of energy $E_e$, given the details of the Ge crystal band structure.
It is computed with the QEdark module~\cite{Essig2016} of the Quantum ESPRESSO package~\cite{QUANTUM-EXPRESSO}. 

For the search of a dark photon, its absorption rate per unit time and target mass is calculated according to~\cite{DarkPhotonAbsorption}:
\begin{equation}
    R=\frac{1}{\rho}\frac{\rho_{DM}}{m_{V}}\kappa^{2}_{\mathrm{eff}}(m_V,\tilde{\sigma})\sigma_1(m_V),
    \label{eq:DP}
\end{equation}
where $m_{V}$ is the dark photon mass and the expected signal is a mono-energetic electron transition of energy $E_e=m_Vc^2$. $\kappa_{\mathrm{eff}}$ is the effective mixing angle which is linearly proportional to the kinetic mixing parameter $\kappa$ between the Standard Model (SM) photon and its hidden counterpart, and $\sigma_1$ is the real part of the complex conductivity $\tilde{\sigma}$. In Ge, the temperature dependence of $\tilde{\sigma}$ above 1~eV is small, allowing us to use the room temperature data from~\cite{DarkPhotonAbsorption} down to 1 eV/c$^2$.
%These effects are stronger below this value, and therefore the dark photon search is not extended below 1 eV/c$^2$.

The signal recorded in the detector, calibrated in eV$_{\mathrm{ee}}$, is $E=(E_e + N\Delta V /\epsilon)/(1+\Delta V /\epsilon)$, thus requiring a discrete distribution function to ascribe a probability $P(N|E_e)$ of producing $N$ electron-hole pairs following an electron transition of energy $E_e$. A variety of ionization models have been proposed~\cite{COMPoisson,Essig2016,CDMS-2019-erratum}. Here, we use the ionization model of~\cite{CDMS-2019-erratum} (with F=0.15) in order to facilitate the comparison of our results with those obtained with this Si phonon-mediated detector.\\

The method to derive 90\%~C.L. upper limits on the DM interaction and absorption rates from the observed spectrum is the same as in~\cite{RED20}. 
%It assumes that the entire rate of events recorded in a given region of interest (ROI) are DM interaction candidates, with no background subtraction.
DM-mass dependent regions of interest (ROIs) are defined on the non-blind sample prior to the unblinding of the search sample. These ROIs are chosen such as to optimize the signal-to-background ratio between the simulated signals and the 30~h of non-blind data.  In order to reduce the importance of statistical fluctuations in the non-blind data set, its energy spectrum was smoothed with a kernel density estimator and a minimum width of 1~eV was imposed to the ROIs. Once these ROIs are defined for the different DM models and mass values, 90\%~C.L. upper limits on their interaction rates are derived using Poisson statistics, considering all events in the search sample ROIs as DM candidates.
%DM-mass dependent regions of interest (ROIs) were defined prior unblinding such as to optimize the expected sensitivity from a KDE background model based on the non-blind data sample.
Because no background subtraction is performed, this procedure yields conservative bounds even if a signal is present in the non-blind data set as it would only result in the ROIs being non optimized.

These limits are shown as the solid red lines in Fig.~\ref{fig:results}.
The top and middle panels are the limits for the interaction cross-section ${\overline{\sigma}}_{e}$ of DM particles with electrons via a heavy ($F_{\mathrm{DM}}=1$) or light mediator ($F_{\mathrm{DM}}\propto 1/q^{2}$), respectively.
The bottom panel shows the limits on the kinetic mixing parameter $\kappa$ of a dark photon with a SM one. Variation on the energy scale of $\pm~2\%$ would affect the limits on $\kappa$ (${\overline{\sigma}}_{e}$) by less than 10\% (20\%).
Temperature effects on $\sigma_1$ in Ge for electron signals above 1~eV$_{\mathrm{ee}}$ are expected to be small: it was nevertheless verified that a 20\% variation of $\sigma_1$ would affect the limits on $\kappa$ by at most $\pm10\%$. 
The excluded event rates are at levels where Earth-shielding effects are negligible~\cite{earthshielding}.

The light red band shows the effect of varying the Fano factor F between 0.30 and the lower bound set by the Bernoulli distribution~\cite{COMPoisson}. The dotted red lines are the limits derived using the linear ionization model described in~\cite{Essig2016}, whereby the $P(N|E_e)$ distribution is replaced by a delta function at the floor value of $N=1+(E_e-E_g)/\epsilon$.
This results in a step-wise evolution of the limits on $\kappa$ as a function of $m_V$, as the sensitivity for a given mass is entirely based on the limit on the rate of $N$-pair events. The noticeable difference around $m_V=3$~eV/$\mathrm{c}^2$ between the dark photon limits obtained when considering these two different ionization models is due to the minimum energy needed to create two-pair events ($\epsilon$ vs. $\epsilon+E_g$).
The EDELWEISS sensitivity below $m_V=3$~eV/$\mathrm{c}^2$ derives from a 90\%~C.L. upper bound of 4 Hz on the efficiency-corrected rate of single-pair events in the detector. The upper limit on the two-pair event rate is 0.08~Hz.
The single-electron rate corresponds to a contribution to the leakage current of the detector of $<6.4\times10^{-19}~\mathrm{A}$.

The present DM constraints extend to much smaller masses than searches based on noble gas detectors~\cite{Xenon10-DP,Darkside} and are competitive with those obtained with Si-based detectors~\cite{SENSEI@MINOS,DAMIC-2019,CDMS-2019-erratum}.
In particular, the present limits are the most stringent ones on the kinetic mixing parameter $\kappa$ for dark photon masses between 6 and 9 eV/$\mathrm{c}^2$.
The better sensitivity of Ge compared to Si for $m_V$~=~1~eV/c$^2$ is due to the difference in gap energies.
In this respect, Ge is a more favorable target for low-mass dark photon searches. For DM-electron scattering above 1 MeV/$\mathrm{c}^2$, Si benefits from more favorable values of $f_c$.

The improvement by an order of magnitude of the detection threshold for electron recoils compared to~\cite{RED20} provides important constraints to understand the origin of the background limiting low-mass DM searches. Further progress in resolving the contributions of heat-only and single-pair events should come from an improvement of the energy resolution.

%Additional information will come from an ongoing investigation of the variation of the shape of the energy spectrum as a function of the applied bias.
%The spectral shape of the DM search data shown in Fig.~\ref{fig:data} suggests a strong contribution of single electron-hole pairs to the sharp rise of events below $10~\mathrm{eV_{ee}}$.
%Still, in order to refute the possible interpretation of this background as either heat-only (HO) or noise-triggered events, an improvement of the resolutions is required. 
In the context of the EDELWEISS-SubGeV program, this will be achieved by upgrading the front-end electronics~\cite{HEMTS} and by operating the NTD sensor at lower temperature to improve its sensitivity. 
To improve the resolution after NTL amplification, the collaboration studies methods to better control the noise induced at large biases
and develops detectors with alternative electrode schemes, such as double-sided vacuum electrodes. 
%Data taking is being pursued to understand the limiting factor of the application of larger biases on the detector (in spite of a relatively low field of 40~V/cm), as well as to investigate change of spectral shape with bias to constrain the contribution to the overall background of HO events that are known to presently limit the experiment sensitivity to low-mass WIMPs. 
The collaboration also investigates sensors based on Superconducting Single-Photon Detectors~\cite{SSPD1,SSPD2} as a possible way to tag ionizing events down to a single charge and thus provide a very efficient rejection of any heat-only background. 
%For the case where the rise at low energy is due to charge leakage, the collaboration investigates different detector sizes and alternate electrode schemes - including double-sided vacuum electrodes - are being developed to achieve electron-hole pair quantization and in order to disentangle possible bulk or surface origin of leakage currents.

In conclusion, the results obtained demonstrate for the first time the high relevance of cryogenic Ge detectors for the search of DM interactions producing eV-scale electron signals and represent an important milestone of the EDELWEISS-SubGeV program which aims at further probing a variety of DM models in the $\mathrm{eV/c^2}$ to $\mathrm{GeV/c^2}$ mass range.
%{\bf Acknowledgments}
\begin{acknowledgments}
The help of the technical staff of the Laboratoire Souterrain de Modane and the participant laboratories is gratefully acknowledged. The EDELWEISS project is supported in part by the German Helmholtz Alliance for Astroparticle Physics (HAP), by the French Agence Nationale pour la Recherche (ANR) and the LabEx Lyon Institute of Origins (ANR-10-LABX-0066) of the Universit\'e de Lyon within the program ``Investissements d'Avenir'' (ANR-11-IDEX-00007), by the P2IO LabEx (ANR-10-LABX-0038) in the framework ``Investissements d'Avenir'' (ANR-11-IDEX-0003-01) managed by the ANR (France), and the Russian Foundation for Basic Research (grant No. 18-02-00159). This project  has  received  funding  from  the  European Union’s Horizon 2020 research and innovation programme under the Marie Skłodowska-Curie Grant Agreement No. 838537.
We thank J.P. Lopez (IP2I), F. Larger, S. Tabtou (Labrador platform IP2I) and the Physics Department of Universit\'{e} Lyon 1 for their contribution to the radioactive sources.
\end{acknowledgments}

%--------Some commands that appear in the biblio file
\newcommand{\arXiv}[1]{\href{https://arxiv.org/abs/#1}{arXiv:#1}}
\newcommand{\oldarXiv}[1]{\href{https://arxiv.org/abs/#1}{#1}}
\newcommand{\DOI}{https://doi.org}

\frenchspacing

\end{document}